\documentclass[prl,preprint,showpacs,showkeys,amsfonts]{revtex4}
\usepackage{graphicx}
\RequirePackage{times}
\RequirePackage{courier}
\RequirePackage{mathptm}

\begin{document}

\title{On the possibility of supercavitation in the quantum ether}
\author{Karl Svozil}
 \email{svozil@tuwien.ac.at}
\homepage{http://tph.tuwien.ac.at/~svozil}
\affiliation{Institut f\"ur Theoretische Physik, University of Technology Vienna,
Wiedner Hauptstra\ss e 8-10/136, A-1040 Vienna, Austria}

\begin{abstract}
In this speculative note some of the conceivable options of supercavitation in the vacuum of quantized field theory are discussed. The resulting modes of propagation could in principle be faster than the velocity of light in unbounded space-time.
\end{abstract}


\maketitle

Supercavitation is a phenomenon by which
an object travels almost frictionless  with high velocities inside a gas bubble
surrounded by a liquid \cite{ashley}.
So far, supercavitation has mostly been applied to military use.

Whereas folklore claims that relativity theory excludes any sort of ether,
the general theory of relativity \cite{einstein-aether,dewitt-synthesis}
and quantum field theory \cite{dirac-aether}
conceptualize entities, often called ``ground state,''  ``vacuum,'' or ``empty space,''
which resemble the pre-relativistic ``mechanic'' ether
in many respects.
Thus it is not totally unreasonable to speculate that
an analogous phenomenon might also occur in the quantum field theoretic vacuum.

The main issues of supercavitation seems to be to
(i) reach and
(ii) maintain
this state, with the additional desideratum of
(iii) flight, in particular direction, control.
Let us therefore first discuss conceivable
options for the formation of ``bubbles of exotic vacuum,''
and then proceed to their maintenance and guidance.

As regards the initial stage of reaching an environment
which might be capable of supercavitating,
one candidate would be the formation of a Casimir vacuum between conducting plates
\cite{milonni-book}, which, due to boundary conditions,
lacks certain modes of the electromagnetic field.
As a consequence, the radiative corrections and the associated renormalized
physical entities such as mass, anomalous magnetic moment
\cite{1996-mass} and the index of refraction  \cite{scharnhorst}
are modified.
The refractive index $n(\omega )<1$ ($\omega$ indicates dependence on photon frequency)
decreases relative to the refractive index in unbounded space (taken to be unity here),
making possible velocities of light $v=c/ n(\omega )>c$ higher than the speed of light in unbounded  vacuum $c$.

Another possibility would be to consider a squeezed vacuum, or an environment
which is characterized by superluminally ``moving'' charge-current patterns;
i.e., synchronized arrays of charges acting similar to phase array radar systems.
Maybe also certain divergences of rotating electromagnetic fields could be utilized
\cite{arda:84}.

The maintenance and steering of supercavitation in the quantum ether remains an open question.
In the case of a cavity in which a  Casimir vacuum is confined
it is unclear if this cavity can be engineered to become a dynamical one.
So far, mostly static problems have been considered; with perfectly conducting, fixed walls.
It may happen, though, that, due to mirror charges or other effects,
the boundary  of this cavity can become a dynamical interface.

It is an often heard belief that superluminal transfer of information
or travel of objects and agents would result in time paradoxes; since then it would be
feasible in certain relativistic coordinate frames to ``reverse the flow of time,''  to
``travel to the past,'' and what not.
Yet, many chronology protection schemes and remedies of different sorts
have been proposed for these scenarios
\cite{godel-sch,recami:01,nahin,svozil-2001-convention},
which  may make such concerns less troublesome.

A {\em caveat}: Let me, instead of a summary, reiterate that this note is very speculative
and contains remotely conceivable options for supercavitation in
the quantum ether. I am aware that this appears to be impossible to realize with
today's scientific and technological means.
Nevertheless, if our ancestors had never dreamed of possibilities of sailing faster
then the wind or rafting quicker than the currents, then we might
not be able to cross the continents, let alone to leave our planet.

%\bibliography{svozil}
%\bibliographystyle{apsrev}
%\bibliographystyle{unsrt}

\end{document}